%% file: phistar_DZero_all_masses.tex
\documentclass[aps,prl,twocolumn,showpacs,superscriptaddress,groupedaddress]{revtex4}  
\usepackage{graphicx}  
\usepackage{dcolumn}   
\usepackage{bm}        
\usepackage{amssymb}   
\usepackage{subfigure}

\def\lumi{10.4}

\def\pythia{{\sc pythia}}
\def\alpgen{{\sc alpgen}}
\def\resbos{{\sc ResBos}}
\def\geant{{\sc geant}}
\def\photos{{\sc photos}}

\newcommand{\mpseudo}{\mbox{$M_{\rm pseudo}$}}
\newcommand{\phiaco}{\mbox{$\phi_{\rm acop}$}}

\newcommand{\at}{\mbox{$a_T$}}
\newcommand{\al}{\mbox{$a_L$}}

\newcommand{\zpt}{\mbox{$p_T^{\ell\ell}$}}

\newcommand{\oneoversigmaphistar}{\mbox{$(1/\sigma)\times (d\sigma/d\phi^{*}_{\eta})$}}
\newcommand{\phistar}{\mbox{$\phi^{*}_{\eta}$}}

\newcommand{\ztt}{\mbox{$Z/\gamma^* \rightarrow \tau^-\tau^+$}}

\newcommand{\met}{\mbox{$E\kern -0.6em/_{\rm T}$}}

\newcommand{\invfb}{fb\mbox{$^{-1}$}}

\begin{document}

\widetext
\voffset=0.75in
\hspace{5.2in} \mbox{FERMILAB-PUB-14-430-E}

\title{Measurement  of the \boldmath{$\phi^*_\eta$} distribution of muon
  pairs with masses
  between \\ 30 and 500~GeV in 10.4~fb\boldmath{$^{-1}$} of \boldmath{$p\bar{p}$} collisions}
\input author_list_phistar.tex

\vskip 0.25cm

\date{29th October 2014.   (Accepted for publication in Phys.~Rev.~D)}

\begin{abstract}
  
  We present a measurement of the distribution of the variable $\phi^*_\eta$ for muon pairs with masses between 30 and 500~GeV, using the complete Run~II data set collected by the D0 detector at the Fermilab Tevatron proton-antiproton collider. This corresponds to an integrated luminosity of 10.4~fb$^{-1}$  at $\sqrt{s}$~=~1.96~TeV. The data are corrected for detector effects and presented in bins of dimuon rapidity and mass. The variable $\phi^*_\eta$ probes the same physical effects as the $Z/\gamma^*$ boson transverse momentum, but is less susceptible to the effects of experimental resolution and efficiency. These are the first measurements at any collider of the $\phi^*_\eta$ distributions for dilepton masses away from the $Z\rightarrow \ell^+\ell^-$ boson mass peak. The data are compared to QCD predictions based on the resummation of multiple soft gluons.

\end{abstract}

\maketitle

\clearpage

Drell-Yan lepton pairs are produced at hadron colliders via quark-antiquark annihilation
and may be produced with a non-zero momentum in the
plane transverse to the beam direction \zpt\ ($\ell = e, \mu, \tau$) due to QCD radiation
from the incoming partons.
Measurements of \zpt\ and related variables in events containing
Drell-Yan lepton pairs thus allow models of initial
state QCD radiation to be tested.
Such models are an important component in the phenomenological interpretation of almost
all experimental measurements and in searches for new physics at hadron colliders.

\begin{figure}[tbph]
    \includegraphics[width=0.45\textwidth]{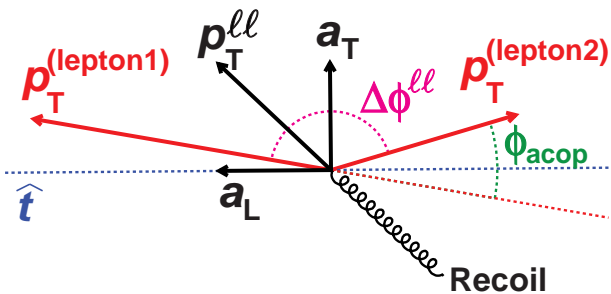}\\
    \caption{(color online) Illustration in the plane transverse to the beam
      direction of the variables defined in the
      text and used to analyse the dilepton transverse
      momentum~\protect\cite{bib:dzero-phistar}.
      The variables
      \at\ and \al~\protect\cite{bib:at} correspond to the decomposition of
      \zpt\ into two orthogonal components relative to the dilepton thrust
      axis $\hat{t}$~\protect\cite{bib:thrust}, as illustrated.
}
    \label{Figure:at}
  \end{figure}

In Ref.~\cite{bib:dzero-phistar} the D0 Collaboration used the variable \phistar~\cite{bib:sa-phistar} 
to study, with unprecedented precision, the \zpt\ distribution of $Z/\gamma^*$ bosons
in dielectron and dimuon final states with dilepton invariant mass
$M_{\ell\ell}$ close to the $Z$ boson pole.
The measurements were presented in bins of dilepton rapidity $|y|$~\cite{bib:rapidity}.
The variable  \phistar\ is defined~\cite{bib:sa-phistar}  as
\begin{equation}
\phistar  \equiv \tan\left(\phiaco/2\right)\sin\theta^*_{\eta},
\label{eqn:phistar}
\end{equation}
where \phiaco\ is the acoplanarity angle, given by
\begin{equation}
\phiaco = \pi - \Delta\phi^{\ell\ell}, 
\end{equation}
and $\Delta\phi^{\ell\ell}$ is the difference in azimuthal angle $\phi$ between the two lepton
candidates.
Figure~\ref{Figure:at} illustrates  relevant
variables in the plane transverse to the beam direction~\cite{bib:dzero-phistar}.
The variable $\theta^*_{\eta}$ is a measure of the scattering angle of the leptons
with respect to the proton beam direction in the rest frame of the dilepton system.
It is defined~\cite{bib:sa-phistar} by
\begin{equation}
\cos(\theta^{*}_{\eta})=\tanh\left[\left(\eta^--\eta^+\right)/2\right], 
\end{equation}
where $\eta^-$ and $\eta^+$ are the pseudorapidities~\cite{bib:pseudo} 
 of the negatively
and positively charged lepton, respectively.
The acoplanarity \phiaco\ gives the degree to which the two leptons deviate from being
back to back in the plane transverse to the beam direction; it is thus
related to \zpt.
Multiplying by $\sin\theta^*_{\eta}$ in Eq.~\ref{eqn:phistar}
corrects for the fact that even for fixed values of $M_{\ell\ell}$
and \zpt, events with different values of $\sin\theta^*_{\eta}$ will correspond to different values of
\phiaco.
The variable \phistar\ is therefore more closely related to \zpt\ than
is \phiaco.
Since \phiaco\ and $\theta^{*}_{\eta}$ depend exclusively on the directions of the two
leptons, which are typically determined with a precision of a milliradian or better, \phistar\ is experimentally very well measured compared
to any quantities, such as \zpt,  that rely on the momenta  of the leptons.

The new experimental variables proposed in Ref.~\cite{bib:sa-phistar} and
exploited by D0 in Ref.~\cite{bib:dzero-phistar}
have prompted a considerable amount of theoretical and experimental  activity.
For example, QCD calculations including the resummation of multiple soft gluon emissions at next-to-next-to-leading log
(NNLL) accuracy~\cite{bib:th0-phistar, bib:th-phistar} and matched to
the next-to-leading order (NLO) Monte Carlo (MC)
calculation MCFM~\cite{bib:MCFM} have been shown
to be consistent with the D0 data~\cite{bib:dzero-phistar} to within the assigned theoretical uncertainties.
The D0 data~\cite{bib:dzero-phistar} have been used as input to
improve the 
predictions from the  \resbos\ MC program~\cite{bib:GNW2013}.
Predictions at NNLL+NLO accuracy for the distribution of \phistar\ of $Z/\gamma^*$
bosons in proton-proton collisions at the
LHC~\cite{bib:lhc-phistar} have been made.
Subsequent experimental measurements of
 the distribution of \phistar\  by the ATLAS
Collaboration~\cite{bib:atlas-phistar} in bins of dilepton rapidity are in agreement with these predictions.
Complementary measurements of the distribution of \phistar\ for
$Z/\gamma^*$ bosons that are highly boosted along the LHC beam
direction have been made by the LHCb
Collaboration~\cite{bib:lhcb-phistar}.
These measurements are also in reasonable agreement with predictions.

In this paper we present measurements of the normalized
\phistar\ distribution of Drell-Yan muon pairs,
\oneoversigmaphistar, in bins of dimuon rapidity in $p\bar{p}$
  collisions at $\sqrt{s}$~=~1.96~TeV.
We update the dimuon measurements
of Ref.~\cite{bib:dzero-phistar} for $70 < M_{\ell\ell} < 110$~GeV
to the complete \lumi~fb$^{-1}$ data set collected by the D0 detector
during Run~2 at the Fermilab Tevatron.
In addition, we extend the
measurements to ``off-peak'' samples of dimuon events and
 consider ranges of  $M_{\ell\ell}$ between 30 and 500~GeV. 
These are the first measurements at any collider of the
\phistar\ distributions of dileptons away from the $Z$ boson mass peak.

As discussed in Ref.~\cite{bib:sa-phistar}, \phistar\ is highly correlated
with the quantity $\at/M_{\ell\ell}$, where the variable
\at~\cite{bib:at} corresponds to one of two orthogonal components
of \zpt\ (as illustrated in Fig.~\ref{Figure:at}).
The width of the \zpt\ distribution is expected to increase
approximately logarithmically with increasing
$M_{\ell\ell}$.
This is because the larger initial-state parton momenta required to produce heavier
dilepton final states allow for harder initial state
radiation~\cite{bib:CSS1985}.
Therefore, the width of the
\phistar\ distribution is expected to decrease with increasing
$M_{\ell\ell}$. 
Measurements of the distribution of \phistar\ over a wide range of
$M_{\ell\ell}$ values allow this prediction to be tested.

Initial state gluon bremsstrahlung represents an important source of
systematic uncertainty in analyses of high mass final states at hadron
colliders, such as
those containing top quarks.
It is therefore interesting to  use the \phistar\ distribution of high mass dilepton final
states to test QCD descriptions of initial state gluon
bremsstrahlung in the relevant mass range.

D0~\cite{bib:DetectorNIM}  is a general purpose detector located at
the Fermilab Tevatron proton-antiproton collider.
The detector has a central-tracking system, consisting of a 
silicon microstrip tracker  and a central fiber tracker, 
both located within a 1.9~T superconducting solenoidal 
magnet, with designs optimized for tracking for $|\eta|<3$.
A liquid-argon and uranium calorimeter has a 
central section covering $|\eta| < 1.1$, and two end calorimeters that extend coverage 
to $|\eta|\approx 4.2$, with all three housed in separate 
cryostats~\cite{bib:run1det}. An outer muon system, at $|\eta|<2$, 
consists of a layer of tracking detectors and scintillation trigger 
counters in front of 1.8~T iron toroidal magnets, followed by two similar layers 
after the toroids~\cite{bib:run2muon}. 

Drell-Yan dimuon MC events are generated with \pythia~\cite{bib:PYTHIA} and passed through a \geant-based~\cite{bib:GEANT}  simulation of the detector.
Backgrounds from \ztt, \mbox{$W\rightarrow \ell\nu$}~(+jets), and $WW\rightarrow \ell\nu\ell\nu$
are simulated using \pythia.
Background from top quark pair production is simulated with
\alpgen~\cite{bib:ALPGEN}, with \pythia\ used for parton showering. 
To simulate the effects of additional
proton-antiproton interactions and detector noise, events
from randomly triggered beam crossings collected during normal data taking are added to the
simulated events.
Background from multijet events is estimated from the data using
samples of events containing poorly isolated muons and same-sign muon pairs.

A second sample of Drell-Yan dimuon MC events (without detector simulation)
is generated using \resbos~\cite{ResBos}. \resbos\ generates $Z$ boson
production with initial state QCD corrections approximately to next-to-next-to-leading order (NNLO) by using 
approximate NNLO Wilson coefficient functions~\cite{bib:GNW2013}, and full NNLL accuracy to account for 
contributions of soft gluon emission~\cite{ResBos, bib:GNW2013}. The $\gamma^{*}$ and $Z/\gamma^{*}$ interference contributions are 
included with initial state QCD corrections to NLO and NNLL accuracy. 
The \resbos\ prediction uses the GNW~\cite{ResBos, bib:GNW2013} non-perturbative function for the
region of 
small \zpt, which is controlled by the parameter $a_Z$.
In our choice of central values and
systematic variations~\cite{bib:CandAZ} for QCD
scales and $a_Z$ we follow Ref.~\cite{bib:GNW2013}.
The CT10 NNLO parton distribution functions (PDFs)~\cite{CTEQ10} are used
and the effects of final state photon
radiation (FSR)  are taken from \photos~\cite{bib:photos}.

In addition to \resbos, we compare our corrected data to the NNLL+NLO predictions
of Ref.~\cite{bib:th-phistar}. 
The NLO PDF sets {\sc CTEQ}6m~\cite{CTEQ6m} are implemented in this
calculation.
The NNLL+NLO predictions of Ref.~\cite{bib:th-phistar} do not include the effects of FSR.
The QCD scales 
are set event by event to the mass of the $Z/\gamma^*$ boson
propagator.

Candidate dimuon events are required to satisfy a
trigger based on the identification of a single muon and to contain
two reconstructed muons.
One of the muons is required to have 
reconstructed track segments in the muon system tracking detectors both inside and outside
the toroidal magnets. The second muon is required to have hits in the muon
system or to have an energy deposit  in the calorimeter that  is
consistent with the passage of a minimum-ionizing particle.
To
ensure an accurate measurement of the muon directions, the two muon
candidates are required to be matched to 
a pair of  particle tracks reconstructed in the
central tracking detectors with momentum transverse to the beam
direction of $p_T$~$>$~15~GeV and
$|\eta|$~$<$~2.
Candidate muons resulting from misidentified hadrons or produced by the decay of hadrons 
are suppressed by requiring that they be isolated from other particles
observed in either the central tracking detectors or the calorimeters.
Requirements are placed on the sum of the $p_T$ of tracks within a
cone of $\Delta R =\sqrt{(\Delta\eta)^2+(\Delta\phi)^2} < 0.5$ around
the muon track and on the sum of
the $E_T$ of calorimeter clusters within an annulus $0.1 < \Delta R <
0.4$ around the muon track.
If more than two muon candidates satisfying the above criteria are
found, the two with the highest $p_T$ are considered.
The muon tracks are required to be oppositely charged.

Contamination from cosmic ray muons is eliminated by requiring that
the muons originate from the  $p\bar{p}$
  collision point on the basis of their impact parameters and times-of-flight, and by rejecting events
in which the two muon candidates 
are back to back in $\eta$ within the experimental resolution.

For $70 < M_{\ell\ell} < 110$~GeV a total of 645k dimuon events is
selected and the total background fraction, arising mainly
from multijet events, is 0.2\%.

Away from the $Z$ boson mass peak it is more difficult to obtain samples of well-measured Drell-Yan dimuon events
with acceptable levels of background, and additional event
selection criteria are imposed.
An important source of contamination in the off-peak samples
arises from Drell-Yan dimuon events that at Born level have a
$Z/\gamma^*$ boson mass outside the selected range in $M_{\ell\ell}$, but are reconstructed with a
value of $M_{\ell\ell}$ within the selected range due to 
FSR or to the mis-measurement of the  $p_T$ of one of the
muon candidates.
We refer below to such events as arising from ``bin migration in $M_{\ell\ell}$''. 
The levels of bin migration in the off-peak signal samples are
estimated and corrected for using the Drell-Yan dimuon MC and are cross checked using control samples in the data. 

Below the $Z$ boson mass peak, dimuon events are selected with $30 < M_{\ell\ell} < 60$~GeV.
To increase the event selection efficiency in this low mass
region and to reduce any kinematic bias on the distribution of
\phistar, the selection criteria are relaxed: we require the leading muon to satisfy
$p_T$~$>$~15~GeV, but allow the second muon  to satisfy
$p_T$~$>$~10~GeV.
The dominant backgrounds in the $30 < M_{\ell\ell} < 60$~GeV sample 
arise  from \ztt\ and bin migration in $M_{\ell\ell}$.
Background from \ztt\ events containing hadronically decaying $\tau$
leptons is suppressed by applying isolation criteria on the muon
candidates that are  more stringent than those used for the $70 <
M_{\ell\ell} < 110$~GeV event sample.
In particular, an additional requirement is placed on the  sum of
the $E_T$ of calorimeter clusters within a cone $\Delta R <
0.1$ around the muon track.
The fraction of the selected event sample arising from \ztt\ background is estimated to be 5.2\%.
The number of selected events that
originate  close to the $Z$ boson mass peak but are reconstructed with
$30 < M_{\ell\ell} < 60$~GeV due to FSR is reduced by excluding
events that contain an isolated photon candidate with $p_T > 15$~GeV.
Bin migration in $M_{\ell\ell}$ from the $Z$ boson mass peak can also arise from events in which the $p_T$ of one of the
muon candidates is underestimated.
This is suppressed using a
``pseudo-mass'' variable, \mpseudo:
the invariant mass of the muon pair is recalculated having set
the magnitude of the $p_T$ of the lower $p_T$ muon to be equal to that of the
higher  $p_T$ muon.
This is under the hypothesis that if an event
originates  close to the $Z$ boson mass peak, but is reconstructed with
$30 < M_{\ell\ell} < 60$~GeV,  the $p_T$ of the lower $p_T$ 
muon candidate has been underestimated.
Events arising from bin migration in $M_{\ell\ell}$ tend to have large values of \mpseudo\ and
candidate events are required to satisfy  $\mpseudo < 75$~GeV.
This requirement rejects only 2\% of Drell-Yan dimuon events with $30 < M_{\ell\ell} < 60$~GeV 
at the generator level.
The fraction of the selected event sample  arising from Drell-Yan
dimuon events  for which the Born level $Z/\gamma^*$ boson propagator
mass is greater than 70~GeV is estimated to be 1.3\%.
Remaining backgrounds amount to 1.6\%  of the selected event sample
and arise mainly from multijet events.
A total of 74k dimuon events is
selected for $30 < M_{\ell\ell} < 60$~GeV.

Above the $Z$ boson mass peak, dimuon events are selected within the two mass
ranges $160 < M_{\ell\ell} < 300$~GeV and $300 < M_{\ell\ell} < 500$~GeV.
In these samples the only significant source of contamination
arises from the moderate resolution in $p_T$ (and thus $M_{\ell\ell}$) in the
compact central tracking detectors of D\O.
The level of bin migration in $M_{\ell\ell}$ in Drell-Yan dimuon events is reduced by imposing tight requirements on the
number of silicon microstrip  and central fiber tracker hits
associated with the muon tracks and the $\chi^2$ of the track fits.
Bin migration in $M_{\ell\ell}$ is further suppressed by rejecting events in which there is a very large asymmetry between the magnitudes of the $p_T$ of the two
muons.
Specifically, it is required that
$$\frac{|\Delta p_T|}{m^2_{\ell\ell}} < 0.004
e^{-M_{\ell\ell}/80} + 0.00115,$$
where $\Delta p_T$ is the difference in the magnitudes of the $p_T$ of the two
muons, and the units of  $\Delta p_T$ and  $M_{\ell\ell}$ are GeV.
For the mass
ranges $160 < M_{\ell\ell} < 300$~GeV and $300 < M_{\ell\ell} < 500$~GeV, respectively, 
the numbers of selected events are 1744 and 207,
and the fractions of the selected event samples arising from bin migration in $M_{\ell\ell}$
are estimated to be 24\% and 44\%.

The observed \phistar\ distributions are corrected for background, and
for experimental efficiency and resolution.
Backgrounds from \ztt, \mbox{$W\rightarrow \ell\nu$}~(+jets),
$WW\rightarrow \ell\nu\ell\nu$,  top quark pairs and multijet events 
are subtracted from the observed \phistar\ distributions.
The corrections to the background-subtracted \phistar\ distribution for
experimental efficiency and resolution (including the effect of
bin migration in $M_{\ell\ell}$) are evaluated
using fully simulated dimuon MC events. 
When evaluating the correction factors, we apply at the MC particle level
the same kinematic
selection criteria on $M_{\ell\ell}$, muon $p_T$, and $|\eta|$ as
in the selection of the data, as specified above.
For this purpose, 
MC particle-level muons are defined after QED final state radiation, which
mimics the measurement of muon momentum in the tracking detector.
In addition, in the low mass dimuon sample ($30 < M_{\ell\ell} < 60$~GeV)
events are rejected if they contain  an FSR photon with transverse energy $E_T^{\gamma}>14$~GeV;
this is in order to  mimic the selection criteria on isolated photons and muon
isolation applied at the detector level.


Since the experimental resolution in \phistar\ is narrower than the
chosen bin widths, the
fractions of accepted events that fall within the same bin in  \phistar\ at the
particle level and reconstructed detector level in the MC are high, having
typical (lowest) values of around 98\% (92\%). 
Therefore, simple bin-by-bin corrections of the \phistar\ distribution
are sufficient.

The fully simulated Drell-Yan dimuon MC events used to calculate the detector corrections are re-weighted at the generator level in two dimensions (\zpt\ and $|y|$) to match the predictions
of \resbos.
In addition, adjustments are made to improve the accuracy of the
following aspects of the detector simulation:  muon $p_T$ scale and resolution,
track $\phi$ and $\eta$ resolutions,
trigger efficiencies, and 
relevant offline reconstruction and selection
efficiencies.
Variations in the above adjustments to the underlying physics and the
detector simulation are included in the assessment of the systematic
uncertainties on the correction factors.
Because of the high bin purities, the systematic uncertainties on the correction factors arising from
variations in the assumed underlying  \phistar\ distribution are found
to be negligible.

The systematic uncertainties due to muon $p_T$ scale
and resolution are small, and arise only due to the kinematic
requirements in the event selection.
The measured \phistar\ distribution is, however, susceptible to
modulations in $\phi$ of the
muon identification and trigger efficiencies, which result, e.g.,  from
detector module boundaries in the  muon system.
Particular care has been taken in the choice of muon identification
criteria in order to minimize such modulations and also to ensure that such modulations are
well simulated in the MC.
For example, the inclusion of muon candidates identified in the
calorimeter reduces the effect of gaps between modules in the outer
muon system.
Nevertheless, accurate modeling of the residual inefficiencies in the
inter-module regions is verified, since this is particularly important in this measurement;
regions of low efficiency that are back-to-back in $\phi$ 
cause the efficiency to modulate as a function of \phistar. 
Accurate modeling of the angular resolution of the central tracking
detectors is another crucial aspect of this analysis. 
The resolution in $\phi$ and $\eta$ is measured in the data using
cosmic ray muons that traverse the detector, since these should
produce events containing two tracks that are exactly back to back
except for the effect of detector resolution.  

Control samples in which one or more of the event selection criteria
are relaxed or inverted are used to test the predicted levels of
background in the off-peak dimuon samples and to assess the associated
systematic uncertainties.
The level of background in the $30 < M_{\ell\ell} < 60$~GeV sample arising from \ztt\ events containing hadronically decaying $\tau$
leptons is verified by checking that the simulation provides a good
description of the sum of
the $E_T$ of calorimeter clusters within  $\Delta R <
0.1$ around the muon track.
In addition, hadrons misidentified as muons are less likely than genuine muons to
be associated with 
reconstructed track segments in the muon system both inside and outside
the toroidal magnets.
 The number and kinematic properties of events containing 
only one such muon candidate, which are enriched in \ztt\ events containing hadronically decaying $\tau$
leptons, are found to be reasonably well described.

In the off-peak samples the predicted levels of  bin migration in $M_{\ell\ell}$ 
are cross checked using control samples.
For $30 < M_{\ell\ell} < 60$~GeV the number and kinematic properties
of the events exclusively rejected by the veto on isolated photons are
well described.
The selection criteria on \mpseudo\ (for $30 < M_{\ell\ell} < 60$~GeV)
and the asymmetry between the magnitudes of the $p_T$ of the two
muons (for $160 < M_{\ell\ell} < 300$~GeV and $300 < M_{\ell\ell} < 500$~GeV) 
introduce a bias on the \phistar\ distributions of the selected event
samples, which has to be accounted for in the MC-derived correction
factors.
It has been verified that the distributions in \mpseudo\ and the $p_T$
asymmetry, having
applied all other selection cuts, are reasonably well described
by the MC.

Systematic uncertainties on the corrections applied to the data arise from
residual uncertainties in 
the modeling of the detector response and the levels of backgrounds.
The following are varied within their uncertainties:
muon $p_T$ scale and resolution;
dependence of trigger and offline identification efficiencies
on $\eta$ and on the proximity to detector module boundaries in
$\phi$.
Systematic uncertainties on the levels of backgrounds are assigned to
cover the statistical uncertainties of the cross checks using control
samples, as well as any residual data-MC discrepancies revealed by these cross checks.
The total experimental systematic uncertainty is evaluated as the quadrature sum of
all the uncertainties discussed above.
In almost all \phistar\ bins the total experimental systematic uncertainty is
substantially smaller than the statistical uncertainty.

The overall QCD uncertainty on the \resbos\ predictions is taken as the quadrature sum of
the changes in the predicted \oneoversigmaphistar\ resulting from 
variations in QCD scales, the non-perturbative parameter $a_Z$ and PDFs.
In our choice of systematic variations~\cite{bib:CandAZ} for QCD
scales and $a_Z$ we follow Ref.~\cite{bib:GNW2013}.
Uncertainties due to PDFs are evaluated using the CT10
NNLO error PDF sets~\protect\cite{CTEQ10}.
In the predictions from \resbos\ the uncertainties arising from QCD
scales are typically a factor of around two larger than those arising
from PDFs or $a_Z$. 
For the NNLL+NLO predictions, the theoretical uncertainties are assessed by variations in renormalization scale, factorization scale and resummation scale between $M_{\ell\ell}/2$ and $2M_{\ell\ell}$, with
the additional requirement that the ratio of any  two of these scales lies between 1/2 and 2~\cite{bib:th-phistar}.

\begin{figure*}[hbtp]
\begin{minipage}[b]{0.97\textwidth}
\centering
\includegraphics[width=0.95\textwidth]{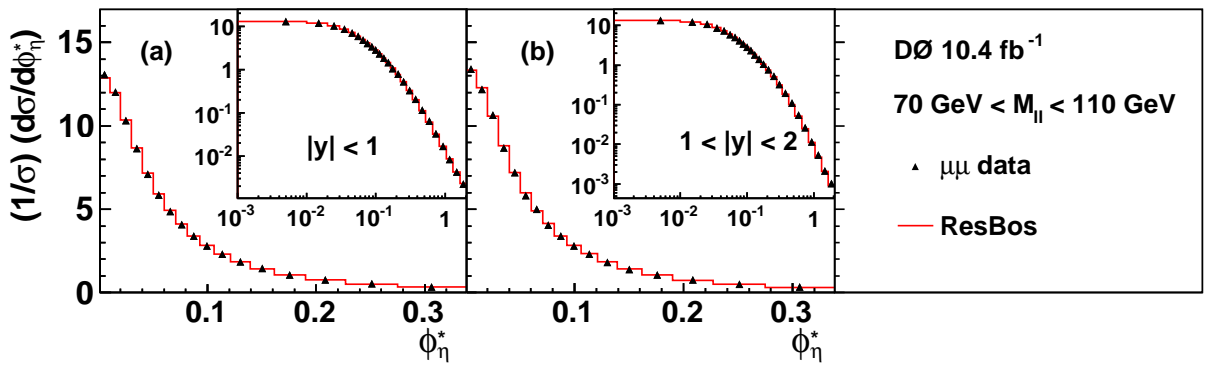}
\caption{(color online) Corrected distributions of
  \oneoversigmaphistar\ in dimuon events with $70 < M_{\ell\ell} <
  110$~GeV for (a)~$|y| < 1$ and
  (b)~$1 < |y| < 2$ in the
  restricted range $0 < \phistar < 0.34$.
 The insets show an extended range of \phistar.
The error bars on the data points represent statistical and systematic
uncertainties combined in quadrature.
The predictions from {\sc ResBos}~\protect\cite{bib:GNW2013} are shown as histograms.
}
\label{fig:peak-phistar}
\end{minipage}
\end{figure*}

Figure~\ref{fig:peak-phistar}
shows the normalized dimuon \phistar\ distributions
\oneoversigmaphistar\ in two bins of dimuon $|y|$ corrected to the particle
level for the kinematic region: 
$70 < M_{\ell\ell} < 110$~GeV, and for both muons $p_T$~$>$~15~GeV and
$|\eta|$~$<$~2.
The data are compared to predictions from
\resbos\ for the same particle-level kinematic region and in the same
bins used for the experimental data.
The values of \oneoversigmaphistar\  are plotted at the center of the
relevant bin in \phistar. 

\begin{figure*}[hbtp]
\begin{minipage}[b]{0.97\textwidth}
\centering
\includegraphics[width=0.95\textwidth]{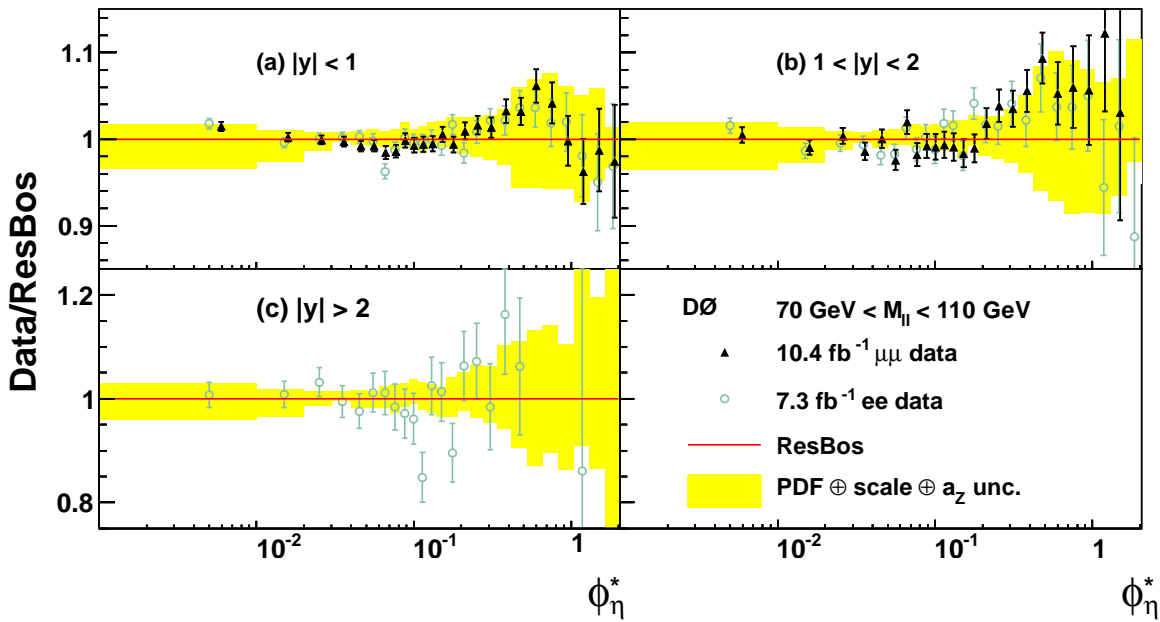}
\caption{(color online) Ratio of the corrected distributions of
  \oneoversigmaphistar\ in dielectron and dimuon data to the
  predictions of
  {\sc ResBos}~\protect\cite{bib:GNW2013}  for $70 < M_{\ell\ell} < 110$~GeV for (a)~$|y| < 1$,
  (b)~$1$~$<$~$|y|$~$<$~2, and
  (c)~$|y|$~$>$~2.
The error bars on the data points represent statistical and systematic
uncertainties combined in quadrature.
The dielectron data are taken
from Ref.~\protect\cite{bib:dzero-phistar} and correspond to an integrated
luminosity of 7.3~fb$^{-1}$.
The band around the \resbos\ prediction represents the quadrature sum of 
uncertainties due to PDFs, QCD scales,  and the non-perturbative  parameter $a_Z$.
}
\label{fig:resbos-peak-phistar}
\end{minipage}
\end{figure*}

Figure~\ref{fig:resbos-peak-phistar} shows the ratio
of the corrected \phistar\ distributions to the \resbos\ predictions for $70 < M_{\ell\ell} < 110$~GeV.
In addition to the dimuon data from the present analysis, the dielectron data
from Ref.~\cite{bib:dzero-phistar}  are shown~\cite{bib:dzero-electrons}. 
Given that the experimental corrections are very different
in the two channels, the consistency of the dielectron and dimuon measurements
represents a powerful cross check of the
corrected distributions.

\begin{figure*}[hbtp]
\begin{minipage}[b]{0.97\textwidth}
\centering
\includegraphics[width=0.95\textwidth]{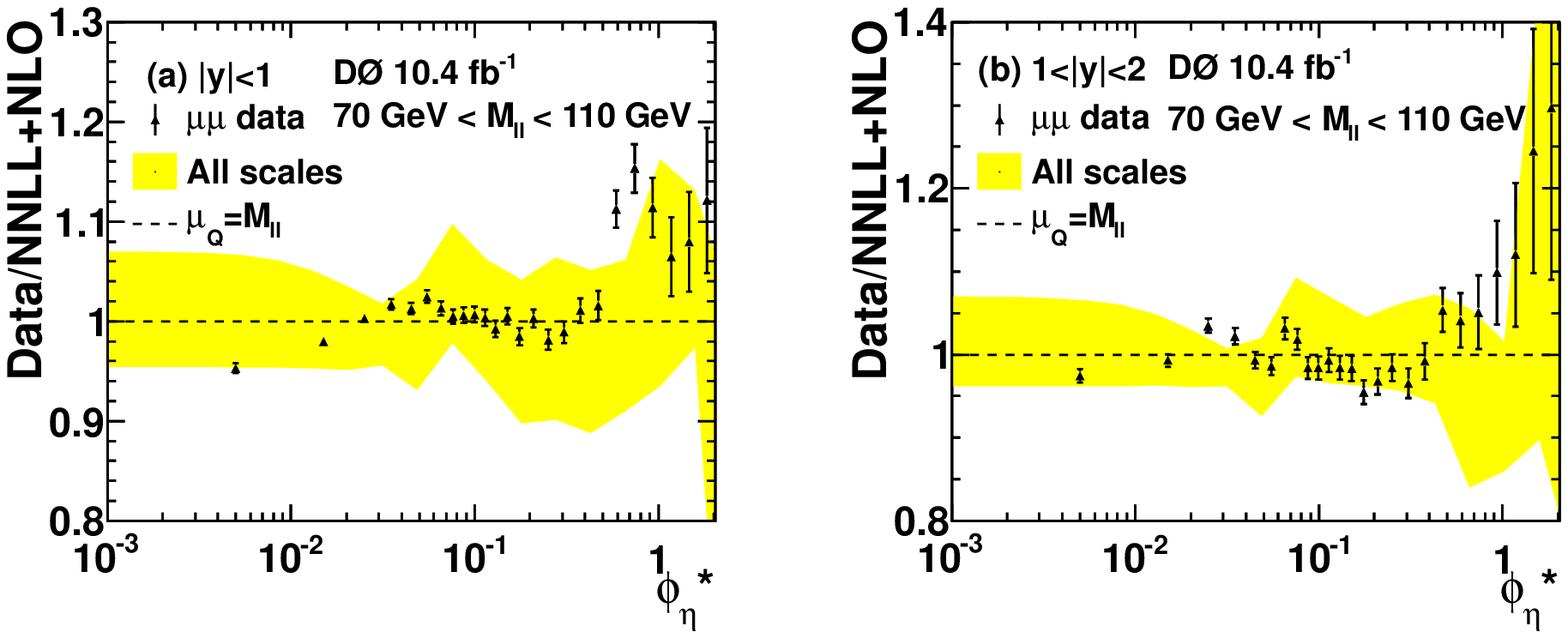}
\caption{(color online) Ratio of the corrected distributions of
  \oneoversigmaphistar\ in  dimuon data to 
  the NNLL+NLO predictions of Ref.~\protect\cite{bib:th-phistar}  for $70 < M_{\ell\ell} < 110$~GeV: (a)~$|y| < 1$ and
  (b)~$1$~$<$~$|y|$~$<$~2.
The error bars on the data points represent statistical and systematic
uncertainties combined in quadrature.
The band around the NNLL+NLO  prediction represents the uncertainty due to variations in the
QCD scales (evaluated by varying the resummation, factorization, and
renormalization scales).
}
\label{fig:NNLL-peak-phistar}
\end{minipage}
\end{figure*}

Figure~\ref{fig:NNLL-peak-phistar} shows  for $70 < M_{\ell\ell} < 110$~GeV the ratio
of the corrected dimuon data to the NNLL+NLO predictions
of Ref.~\cite{bib:th-phistar}. 

\begin{figure*}[hbtp]
\begin{minipage}[b]{0.97\textwidth}
\centering
 \subfigure{\includegraphics[width=0.49\textwidth]{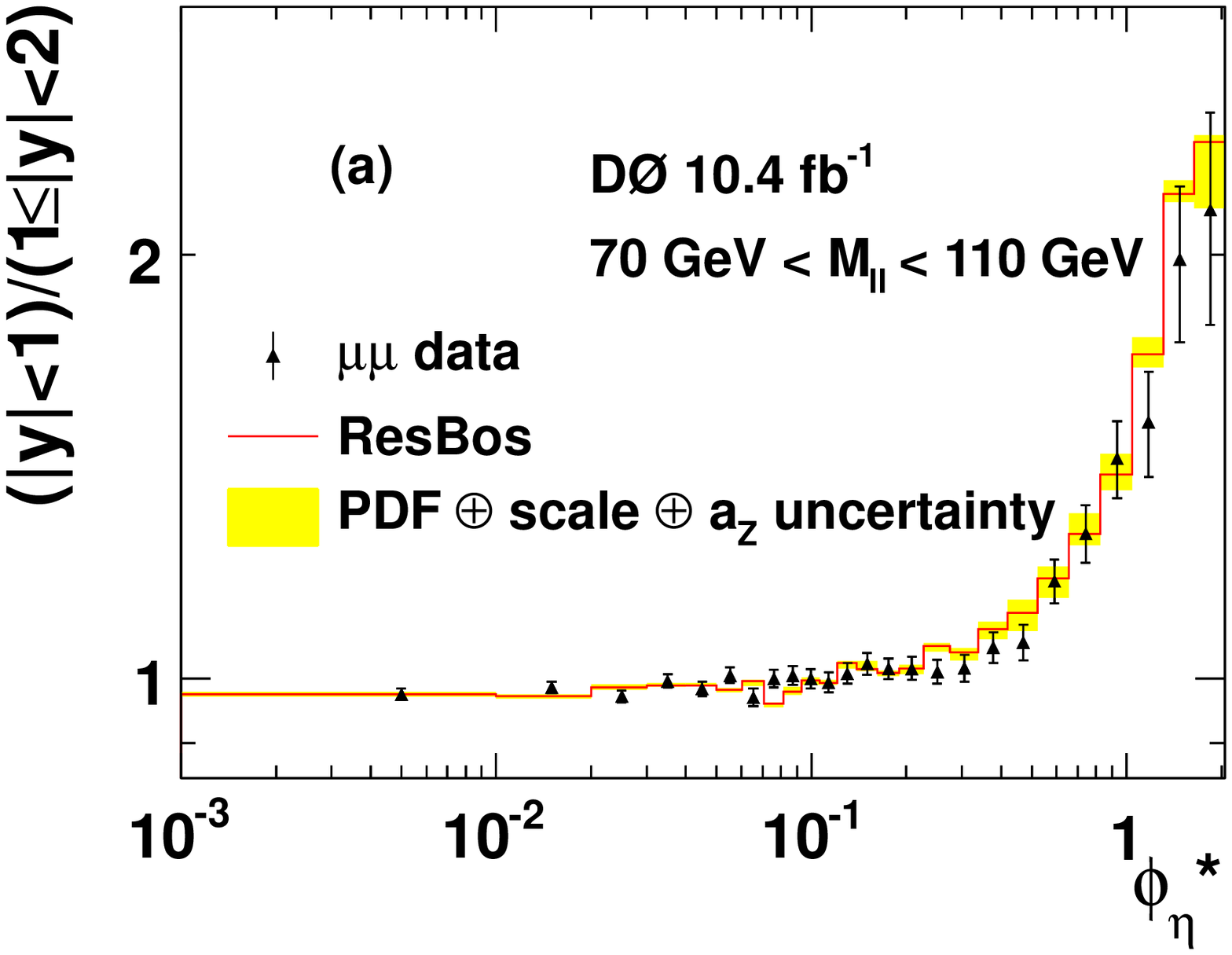}}
  \subfigure{\includegraphics[width=0.49\textwidth]{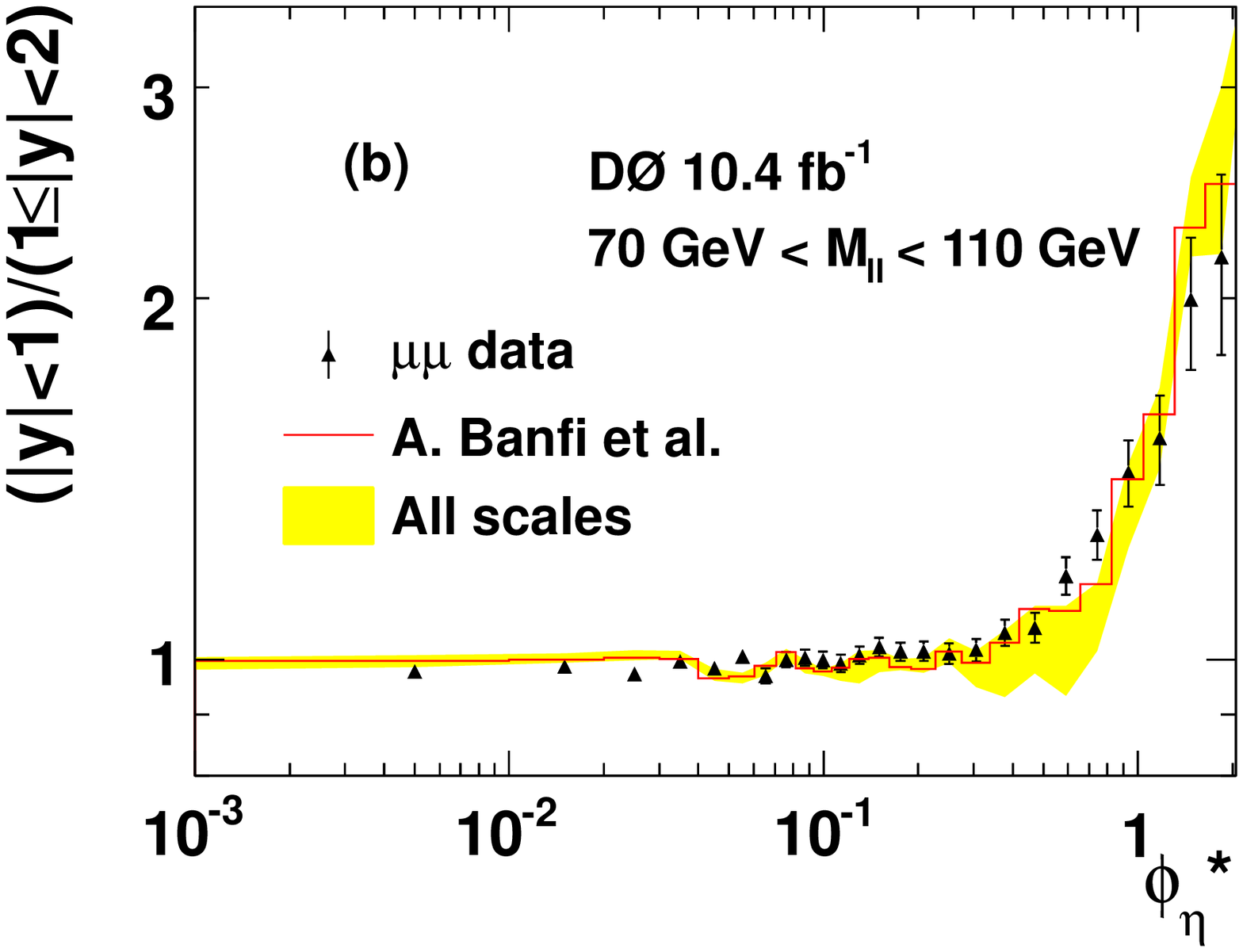}}\\
\caption{(color online)  Ratio of the $(1/\sigma)(d\sigma/d\phistar)$ distribution
 in the central rapidity region ($|y|<1$) to that in the forward 
rapidity region ($1<|y|<2$). 
The corrected dimuon data are compared to the predictions from
 (a)~{\sc ResBos}~\protect\cite{bib:GNW2013}  and from (b)~NNLL+NLO~\protect\cite{bib:th-phistar}.  
The error bars on the data points represent statistical and systematic
uncertainties combined in quadrature, assuming no correlation
between the two rapidity regions. 
The theoretical predictions  are represented by histograms and the
band shows the theoretical uncertainties, taking  correlations
between the two rapidity regions into account.
}
\label{fig:NNLL-peak-phistar_ratio}
\end{minipage}
\end{figure*}

Figures~\ref{fig:resbos-peak-phistar}
and~\ref{fig:NNLL-peak-phistar} show that the theoretical uncertainties 
arising from QCD scale variations and PDFs are
large compared to the experimental uncertainties.
Within the quoted uncertainties both predictions are consistent with
the corrected data.
Figure~\ref{fig:NNLL-peak-phistar_ratio}
 shows the ratio of the $(1/\sigma)(d\sigma/d\phistar)$ distribution
 in the central rapidity region ($|y|<1$) to that in the forward 
rapidity region ($1<|y|<2$). 
The corrected dimuon data are compared to the predictions from
{\sc ResBos}~\cite{bib:GNW2013}  and from the NNLL+NLO calculations~\cite{bib:th-phistar}. 
Figure~\ref{fig:NNLL-peak-phistar_ratio}
 shows that the theoretical uncertainties largely  cancel in
this ratio and that the predictions are consistent with the data.

\begin{figure*}[hbtp]
\begin{minipage}[b]{0.97\textwidth}
\centering
\includegraphics[width=0.95\textwidth]{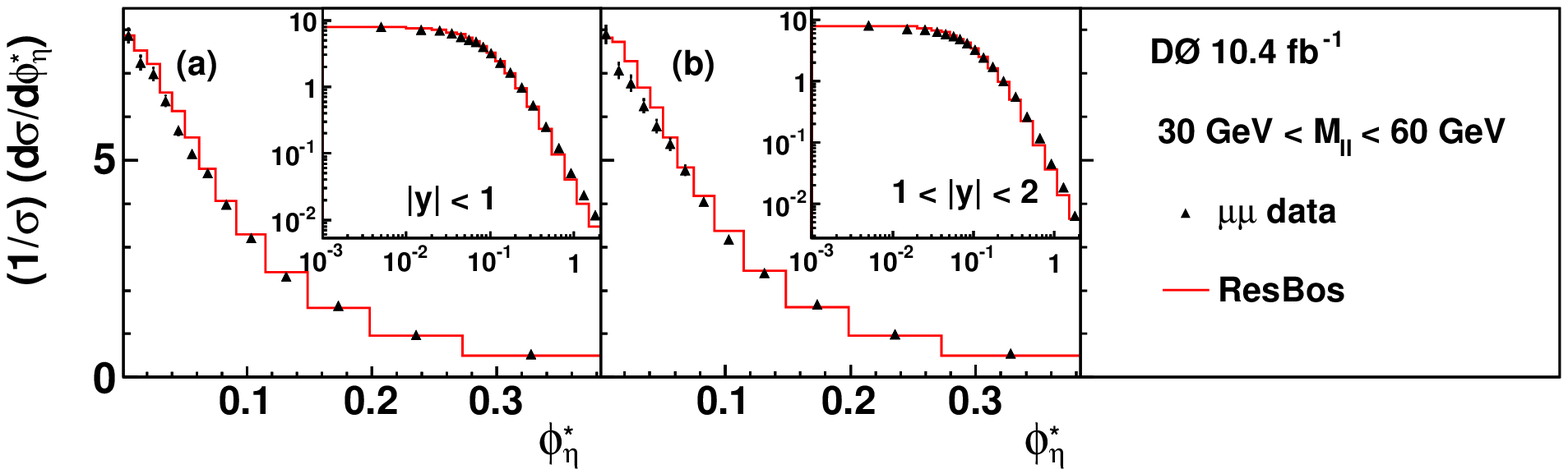}
\caption{(color online) Corrected distributions of
  \oneoversigmaphistar\ in dimuon events with  $30 < M_{\ell\ell} <
  60$~GeV for (a)~$|y| < 1$ and
  (b)~$1 < |y| < 2$ in the
  restricted range $0 < \phistar < 0.38$. 
 The insets show an extended range of \phistar.
The error bars on the data points represent statistical and systematic
uncertainties combined in quadrature.
The predictions from {\sc ResBos}~\protect\cite{bib:GNW2013} are shown as histograms.
}
\label{fig:low-phistar}
\end{minipage}
\end{figure*}

Figure~\ref{fig:low-phistar}
shows the normalized dimuon \phistar\ distributions
\oneoversigmaphistar\ in two bins of  $|y|$, corrected to the particle level with
kinematic requirements: 
$30 < M_{\ell\ell} < 60$~GeV, and for both muons 
$|\eta|$~$<$~2.
The leading muon is required to satisfy
$p_T$~$>$~15~GeV and the second muon  is required to satisfy
$p_T$~$>$~10~GeV.
In addition, events are required at particle level to contain no
FSR photon with transverse energy $E_T^{\gamma}>14$~GeV.
The corrected data are compared to predictions from
{\sc ResBos}~\protect\cite{bib:GNW2013} with the same particle-level kinematic cuts applied.

\begin{figure*}[hbtp]
\begin{minipage}[b]{0.97\textwidth}
\centering
\includegraphics[width=0.95\textwidth]{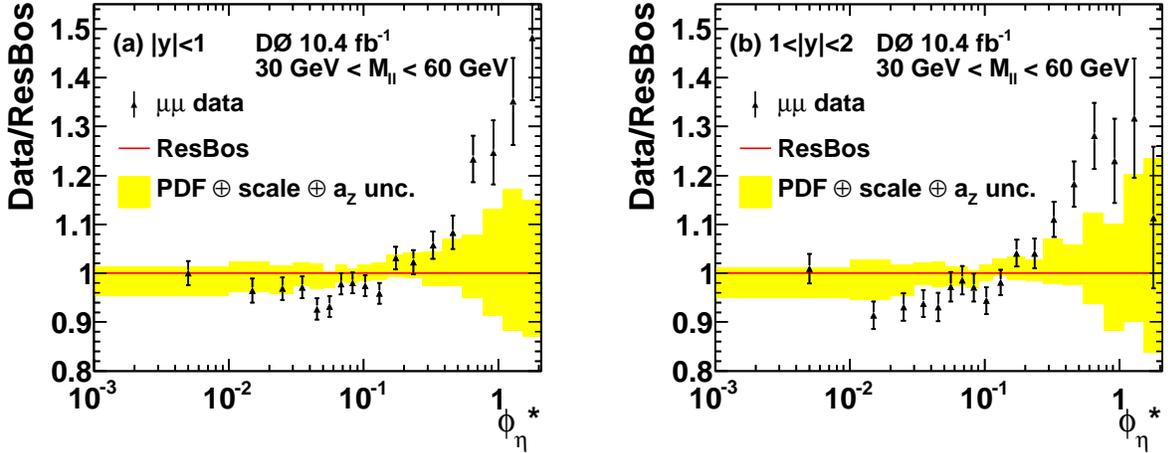}
\caption{(color online) Ratio of the corrected distributions of
  \oneoversigmaphistar\ to 
  {\sc ResBos}~\protect\cite{bib:GNW2013}   in dimuon events with  $30 < M_{\ell\ell} <
  60$~GeV for (a)~$|y| < 1$ and
  (b)~$1 < |y| < 2$.
Statistical and systematic
uncertainties are combined in quadrature.
The band around the \resbos\ prediction represents the quadrature sum of 
uncertainties due to PDFs, QCD scales,  and the non-perturbative  parameter $a_Z$.
}
\label{fig:resbos-low-phistar}
\end{minipage}
\end{figure*}

\begin{figure*}[hbtp]
\begin{minipage}[b]{0.97\textwidth}
\centering
\includegraphics[width=0.95\textwidth]{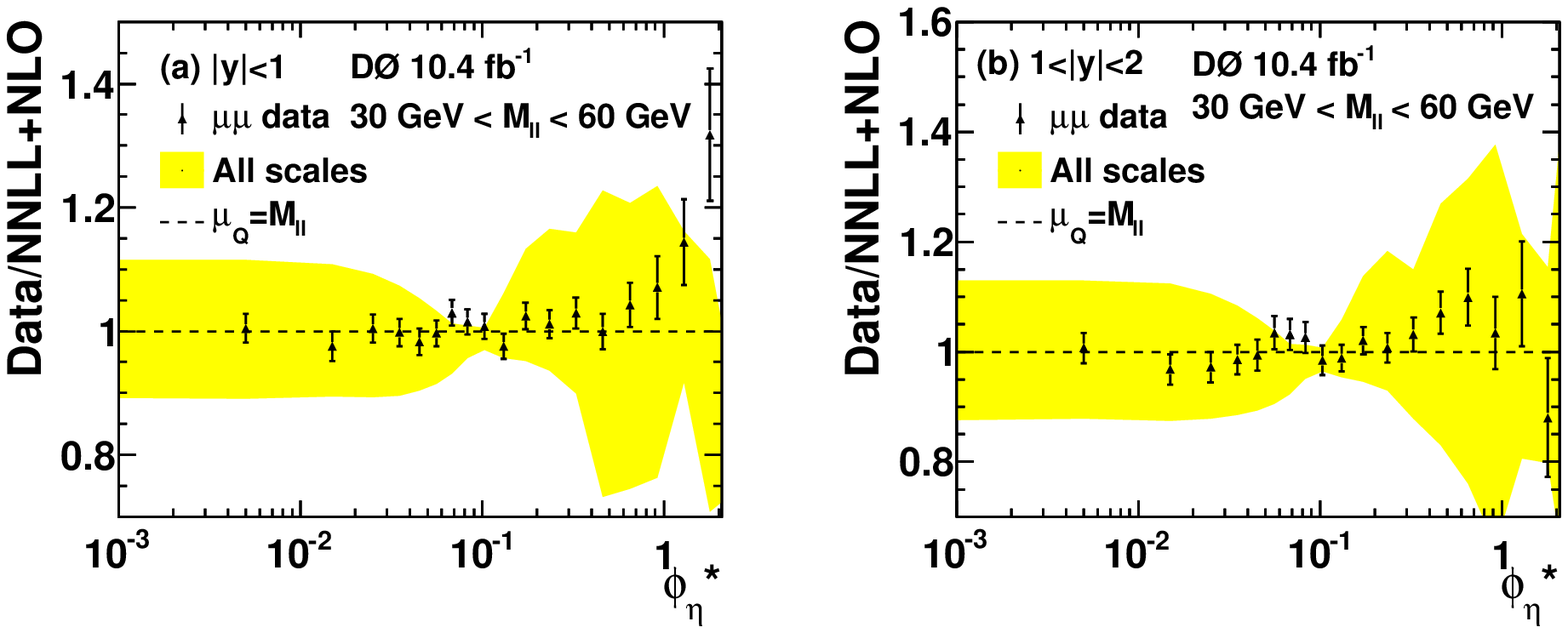}
\caption{(color online) Ratio of the corrected distributions of
  \oneoversigmaphistar\ to 
  the NNLL+NLO predictions of Ref.~\protect\cite{bib:th-phistar, bib:Tomlinson}  in dimuon events with  $30 < M_{\ell\ell} <
  60$~GeV for (a)~$|y| < 1$ and
  (b)~$1 < |y| < 2$.
Statistical and systematic uncertainties are combined in quadrature. 
The band around the NNLL+NLO  prediction represents the uncertainty due to variations in the
QCD scales (evaluated by varying the resummation, factorization, and
renormalization scales).
}
\label{fig:NNLL-low-phistar}
\end{minipage}
\end{figure*}

Figure~\ref{fig:resbos-low-phistar} shows the ratio
of the corrected \phistar\ distributions to the \resbos\ predictions for $30 < M_{\ell\ell} < 60$~GeV.
Figure~\ref{fig:NNLL-low-phistar} shows  the ratio
of the same data to the NNLL+NLO predictions of Ref.~\cite{bib:th-phistar, bib:Tomlinson}.
At high values of \phistar\ the  prediction from \resbos\ agrees
less well with data than is the case  in the region of the $Z$ boson mass peak. 
A known deficiency of the \resbos\ prediction for $\phistar>0.5$ in the low
mass region is the absence of the NNLO correction factor for the
photon exchange diagram.

\begin{figure*}[hbtp]
\begin{minipage}[b]{0.97\textwidth}
\centering
\includegraphics[width=0.996\textwidth]{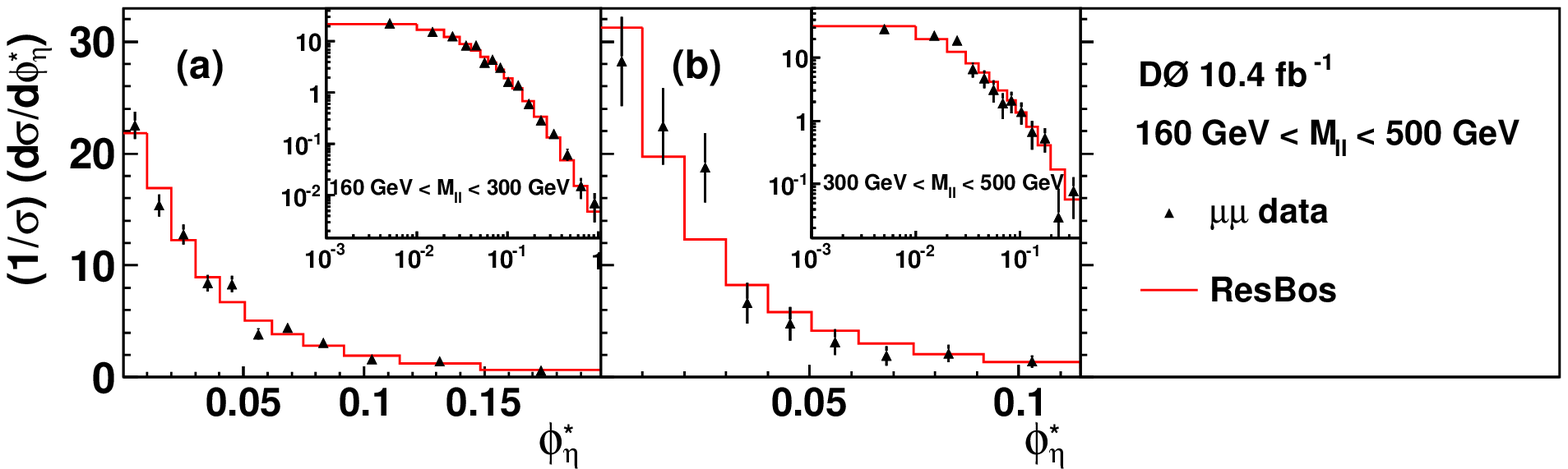}
\caption{(color online) Corrected distributions of
  \oneoversigmaphistar\ for dimuon events with (a)~$160 < M_{\ell\ell}
  < 300$~GeV and (b)~$300 < M_{\ell\ell} < 500$~GeV for a
  restricted range of \phistar. 
 The insets show an extended range of \phistar.
The error bars on the data points represent statistical and systematic
uncertainties combined in quadrature.
The predictions from {\sc ResBos}~\protect\cite{bib:GNW2013} are shown as the red histogram.
}
\label{fig:high-phistar}
\end{minipage}
\end{figure*}

\begin{figure*}[hbtp]
\begin{minipage}[b]{0.97\textwidth}
\centering
\includegraphics[width=0.996\textwidth]{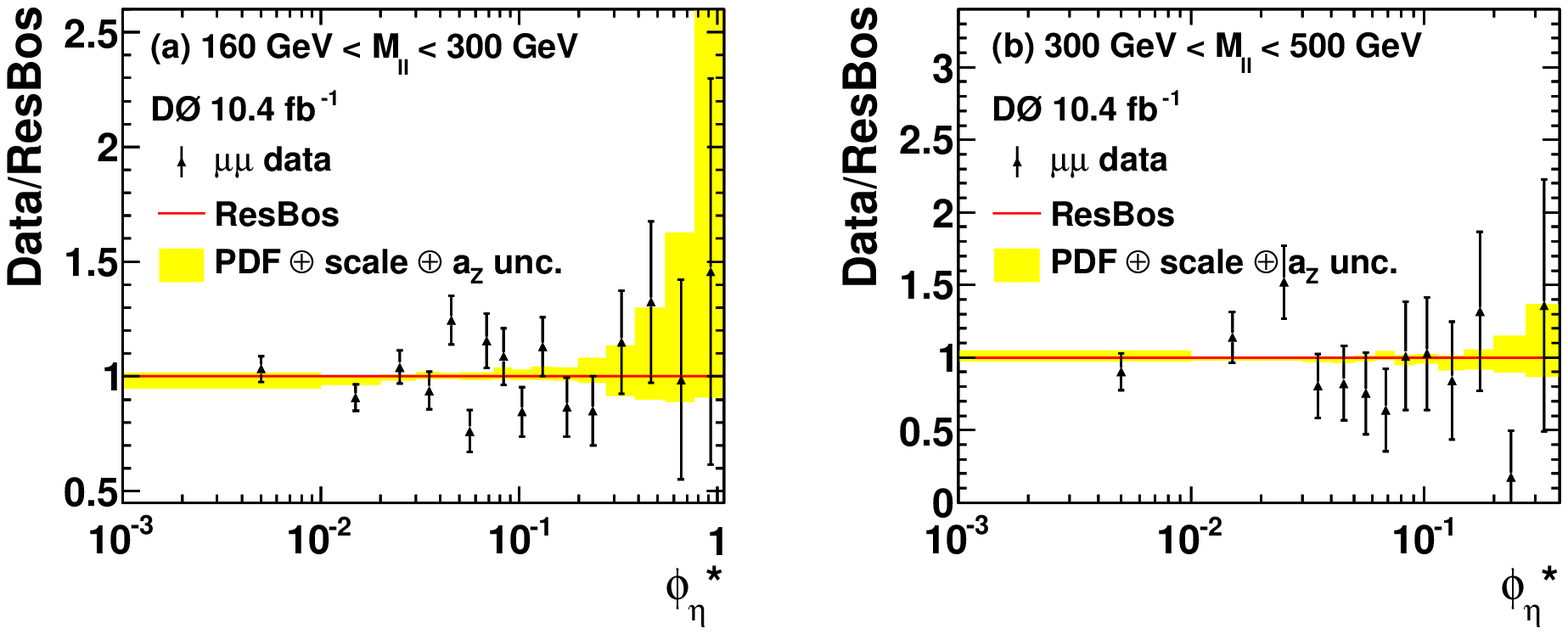}
\caption{(color online) Ratio of the corrected distributions of
  \oneoversigmaphistar\ to 
   {\sc ResBos}~\protect\cite{bib:GNW2013}  for (a)~$160 < M_{\ell\ell} < 300$~GeV and (b)~$300 < M_{\ell\ell} < 500$~GeV.
Statistical and systematic
uncertainties are combined in quadrature.
The band around the \resbos\ prediction represents the quadrature sum of 
uncertainties due to PDFs, QCD scales,  and the non-perturbative  parameter $a_Z$.
}
\label{fig:resbos-high-phistar}
\end{minipage}
\end{figure*}

Figure~\ref{fig:high-phistar}
shows the normalized dimuon \phistar\ distributions
\oneoversigmaphistar, corrected to the particle level with
kinematic requirements: 
$160 < M_{\ell\ell} < 300$~GeV and $300 < M_{\ell\ell} < 500$~GeV, and for both muons $p_T$~$>$~15~GeV and
$|\eta|$~$<$~2.
The data are compared to predictions from
\resbos\ with the same particle-level kinematic  requirements applied.
Figure~\ref{fig:resbos-high-phistar} shows the ratios
of the corrected \phistar\ distributions to the \resbos\ predictions.
Within the fairly large statistical uncertainties, the predictions are
consistent with the corrected data.
As pointed out above there is no NNLO correction factor for the
photon exchange diagram in \resbos.


\begin{figure*}[hbtp]
\begin{minipage}[b]{0.97\textwidth}
\centering
\includegraphics[width=0.95\textwidth]{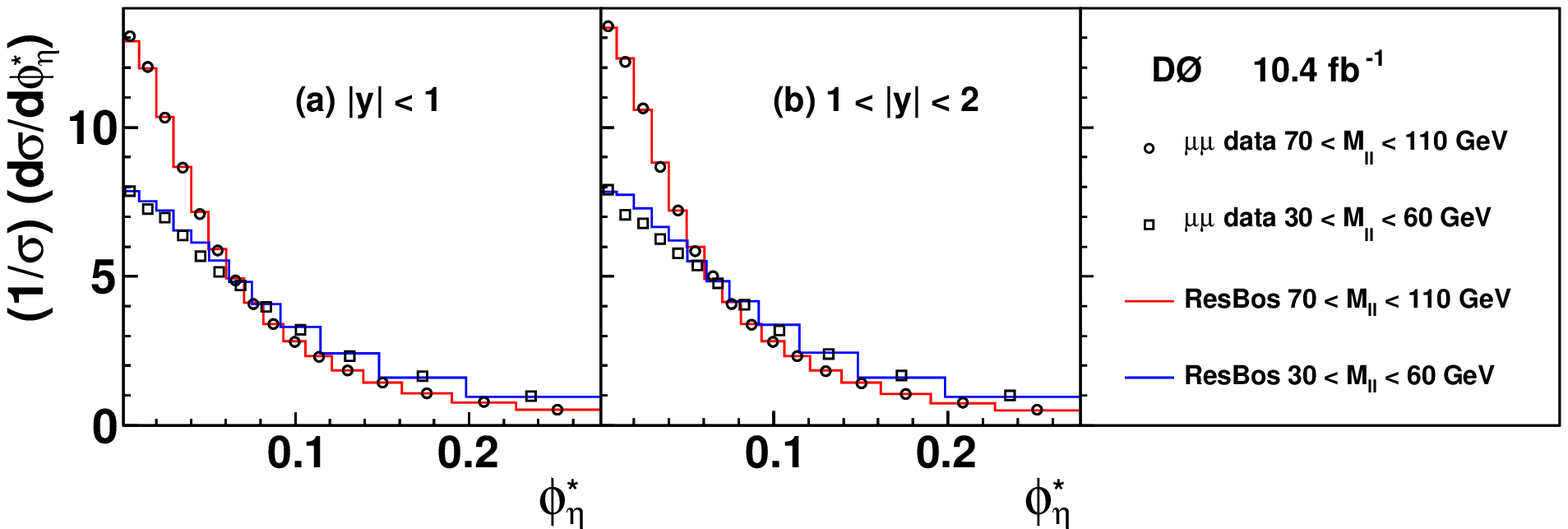}
\caption{(color online) Comparison of corrected distributions of
  \oneoversigmaphistar\ in dimuon events with  $30 < M_{\ell\ell} <
  60$~GeV and  $70 < M_{\ell\ell} < 130$~GeV for (a)~$|y| < 1$ and
  (b)~$1 < |y| < 2$ in the
  restricted range $0 < \phistar < 0.28$.
The error bars on the data points represent statistical and systematic
uncertainties combined in quadrature.
The predictions from {\sc ResBos}~\protect\cite{bib:GNW2013} are shown as histograms.
}
\label{fig:peaklow-phistar}
\end{minipage}
\end{figure*}

The corrected distributions of
  \oneoversigmaphistar\ in the two dimuon mass ranges  $30 < M_{\ell\ell} <
  60$~GeV and  $70 < M_{\ell\ell} < 130$~GeV are compared in Fig.~\ref{fig:peaklow-phistar}.
As discussed above, the width of the
\phistar\ distribution is expected to decrease with increasing
$M_{\ell\ell}$. 
Fig.~\ref{fig:peaklow-phistar}
 shows that the data are consistent
with this expectation and that \resbos\ provides a good
description of this behavior.
The numbers of selected events in the dimuon mass ranges
$160 < M_{\ell\ell} < 300$~GeV and $300 < M_{\ell\ell} < 500$~GeV are
insufficient to allow us to present the distributions of
  \oneoversigmaphistar\ in the two separate ranges of $|y|$ shown in
  Fig.~\ref{fig:peaklow-phistar}.
However, the dependence on $|y|$ is small and a comparison between 
Figs.~\ref{fig:high-phistar} and~\ref{fig:peaklow-phistar}
 shows that the distributions of
  \oneoversigmaphistar\ continue to become more narrow with increasing
  dimuon mass in the region above the $Z$ boson mass peak.

In summary, using \lumi~\invfb\ of  $p\bar{p}$
  collisions 
we  have measured the normalized \phistar\ distribution
\oneoversigmaphistar\ in two bins of dimuon rapidity
and four bins of dimuon mass.
Relative to the results presented in Ref.~\cite{bib:dzero-phistar},
these measurements in the dimuon channel represent an extension to the
full D0 data set and also to regions of dimuon mass away from the $Z$
boson mass peak. 
The data are well described within the theoretical
uncertainties by  the \resbos\ MC and  by the predictions at NNLL+NLO
accuracy of Ref.~\cite{bib:th-phistar, bib:Tomlinson}.
In the $Z$ boson mass peak region, $70 < M_{\ell\ell} < 110$~GeV, 
the theoretical uncertainties 
shown in Figs.~\ref{fig:resbos-peak-phistar}
and~\ref{fig:NNLL-peak-phistar} are
large compared to the experimental uncertainties.
Figure~\ref{fig:NNLL-peak-phistar_ratio} shows
the ratio of the $(1/\sigma)(d\sigma/d\phistar)$ distribution
 in the central rapidity region ($|y|<1$) to that in the forward 
rapidity region ($1<|y|<2$). 
The  theoretical uncertainties largely  cancel in
this ratio and the QCD predictions are consistent with the data.
The data are consistent
with the expectation that the width of the
\phistar\ distribution  decreases with increasing
$M_{\ell\ell}$.
The measurements of \phistar\ distributions above the $Z$ boson mass peak may help constrain 
systematic uncertainties arising from initial state gluon
bremsstrahlung in analyses of other high mass final states, such as
those containing top quarks.

Tables of corrected \oneoversigmaphistar\ distributions
for each $|y|$ bin and range of $M_{\ell\ell}$ are provided in the
appendix.
In some of these tables results are given for a larger range of
\phistar\ than is shown in the corresponding figures. 

We thank the authors of Refs.~\cite{bib:GNW2013} and~\cite{bib:th-phistar},
in particular Marco Guzzi and Lee Tomlinson, repectively, for their help in evaluating
predictions to be compared with the new off-peak measurements.
We thank the staffs at Fermilab and collaborating institutions,
and acknowledge support from the
Department of Energy and National Science Foundation (United States of America);
Alternative Energies and Atomic Energy Commission and
National Center for Scientific Research/National Institute of Nuclear and Particle Physics  (France);
Ministry of Education and Science of the Russian Federation, 
National Research Center ``Kurchatov Institute" of the Russian Federation, and 
Russian Foundation for Basic Research  (Russia);
National Council for the Development of Science and Technology and
Carlos Chagas Filho Foundation for the Support of Research in the State of Rio de Janeiro (Brazil);
Department of Atomic Energy and Department of Science and Technology (India);
Administrative Department of Science, Technology and Innovation (Colombia);
National Council of Science and Technology (Mexico);
National Research Foundation of Korea (Korea);
Foundation for Fundamental Research on Matter (The Netherlands);
Science and Technology Facilities Council and The Royal Society (United Kingdom);
Ministry of Education, Youth and Sports (Czech Republic);
Bundesministerium f\"{u}r Bildung und Forschung (Federal Ministry of Education and Research) and 
Deutsche Forschungsgemeinschaft (German Research Foundation) (Germany);
Science Foundation Ireland (Ireland);
Swedish Research Council (Sweden);
China Academy of Sciences and National Natural Science Foundation of China (China);
and
Ministry of Education and Science of Ukraine (Ukraine).

\clearpage
\newpage

\appendix

\noindent {\bf\Large Appendix: Tables of results}

 \begin{table}[hbpt]
  \begin{minipage}[b]{0.97\textwidth}
       \caption{Table of results for the dimuon channel for
         $|y|<1$ region with $70 < M_{\ell\ell} < 110$~GeV. The first quoted uncertainty is statistical and the second is the
total experimental systematic uncertainty.   
}
   \centering
   \begin{tabular}{|c|c|c|}
      \hline \hline
   bin & $\phistar$ range & $1/\sigma$~$d\sigma/d\phistar$ \\
   \hline

\input{Table_peak_1.tex}

   \hline \hline
   \end{tabular}
   \label{Table:results_dimu_y1corr_effsystematics}
    \end{minipage}
     \end{table}

   \begin{table}[hbpt]
   \begin{minipage}[b]{0.97\textwidth}
       \caption{Table of results for the dimuon channel for $1<|y|<2$ region with $70 < M_{\ell\ell} < 110$~GeV. The first quoted uncertainty is statistical and the second is the
total experimental systematic uncertainty.}
   \centering
   \begin{tabular}{|c|c|c|}
      \hline \hline
   bin & $\phistar$ range & $1/\sigma$~$d\sigma/d\phistar$ \\
   \hline

\input{Table_peak_2.tex}

   \hline \hline
   \end{tabular}
   \label{Table:results_dimu_y2corr_effsystematics}
   \end{minipage}
   \end{table}

  \begin{table}[hbpt]
  \begin{minipage}[b]{0.97\textwidth}
       \caption{Table of results for the dimuon channel for $|y|<1$ region  $30 < M_{\ell\ell} < 60$~GeV. The first quoted uncertainty is statistical and the second is the
total experimental systematic uncertainty.}
   \centering
   \begin{tabular}{|c|c|c|}
      \hline \hline
   bin & $\phistar$ range & $1/\sigma$~$d\sigma/d\phistar$ \\
   \hline
\input{Table_low_1.tex}

   \hline \hline
   \end{tabular}
   \label{Table:results_diem_y1low}
    \end{minipage}
     \end{table}

   \begin{table}[hbpt]
     \begin{minipage}[b]{0.97\textwidth}
       \caption{Table of results for the dimuon channel for $1<|y|<2$ region  $30 < M_{\ell\ell} < 60$~GeV. The first quoted uncertainty is statistical and the second is the
total experimental systematic uncertainty.}
   \centering
   \begin{tabular}{|c|c|c|}
      \hline \hline
   bin & $\phistar$ range & $1/\sigma$~$d\sigma/d\phistar$ \\
   \hline
\input{Table_low_2.tex}

   \hline \hline
   \end{tabular}
   \label{Table:results_diem_y2low}
   \end{minipage}
   \end{table}

 \begin{table}[hbpt]
 \begin{minipage}[b]{0.97\textwidth}
       \caption{Table of results for the dimuon channel for $160 < M_{\ell\ell} < 300$~GeV region. The first quoted uncertainty is statistical and the second is the
total experimental systematic uncertainty.}
   \centering
   \begin{tabular}{|c|c|c|}
      \hline \hline
   bin & $\phistar$ range & $1/\sigma$~$d\sigma/d\phistar$ \\
   \hline
\input{high/result_0.tex}

   \hline \hline
   \end{tabular}
   \label{Table:results_diem_y1high}
    \end{minipage}
     \end{table}

   \begin{table}[hbpt]
   \begin{minipage}[b]{0.97\textwidth}
       \caption{Table of results for the dimuon channel for $300 < M_{\ell\ell} < 500$~GeV region. The first quoted uncertainty is statistical and the second is the
total experimental systematic uncertainty.}
   \centering
   \begin{tabular}{|c|c|c|}
      \hline \hline
   bin & $\phistar$ range & $1/\sigma$~$d\sigma/d\phistar$ \\
   \hline
\input{high/result_3.tex}

   \hline \hline
   \end{tabular}
   \label{Table:results_diem_y2high}
   \end{minipage}
   \end{table}

\end{document}

%% file: author_list_phistar.tex
\affiliation{LAFEX, Centro Brasileiro de Pesquisas F\'{i}sicas, Rio de Janeiro, Brazil}
\affiliation{Universidade do Estado do Rio de Janeiro, Rio de Janeiro, Brazil}
\affiliation{Universidade Federal do ABC, Santo Andr\'e, Brazil}
\affiliation{University of Science and Technology of China, Hefei, People's Republic of China}
\affiliation{Universidad de los Andes, Bogot\'a, Colombia}
\affiliation{Charles University, Faculty of Mathematics and Physics, Center for Particle Physics, Prague, Czech Republic}
\affiliation{Czech Technical University in Prague, Prague, Czech Republic}
\affiliation{Institute of Physics, Academy of Sciences of the Czech Republic, Prague, Czech Republic}
\affiliation{Universidad San Francisco de Quito, Quito, Ecuador}
\affiliation{LPC, Universit\'e Blaise Pascal, CNRS/IN2P3, Clermont, France}
\affiliation{LPSC, Universit\'e Joseph Fourier Grenoble 1, CNRS/IN2P3, Institut National Polytechnique de Grenoble, Grenoble, France}
\affiliation{CPPM, Aix-Marseille Universit\'e, CNRS/IN2P3, Marseille, France}
\affiliation{LAL, Universit\'e Paris-Sud, CNRS/IN2P3, Orsay, France}
\affiliation{LPNHE, Universit\'es Paris VI and VII, CNRS/IN2P3, Paris, France}
\affiliation{CEA, Irfu, SPP, Saclay, France}
\affiliation{IPHC, Universit\'e de Strasbourg, CNRS/IN2P3, Strasbourg, France}
\affiliation{IPNL, Universit\'e Lyon 1, CNRS/IN2P3, Villeurbanne, France and Universit\'e de Lyon, Lyon, France}
\affiliation{III. Physikalisches Institut A, RWTH Aachen University, Aachen, Germany}
\affiliation{Physikalisches Institut, Universit\"at Freiburg, Freiburg, Germany}
\affiliation{II. Physikalisches Institut, Georg-August-Universit\"at G\"ottingen, G\"ottingen, Germany}
\affiliation{Institut f\"ur Physik, Universit\"at Mainz, Mainz, Germany}
\affiliation{Ludwig-Maximilians-Universit\"at M\"unchen, M\"unchen, Germany}
\affiliation{Panjab University, Chandigarh, India}
\affiliation{Delhi University, Delhi, India}
\affiliation{Tata Institute of Fundamental Research, Mumbai, India}
\affiliation{University College Dublin, Dublin, Ireland}
\affiliation{Korea Detector Laboratory, Korea University, Seoul, Korea}
\affiliation{CINVESTAV, Mexico City, Mexico}
\affiliation{Nikhef, Science Park, Amsterdam, the Netherlands}
\affiliation{Radboud University Nijmegen, Nijmegen, the Netherlands}
\affiliation{Joint Institute for Nuclear Research, Dubna, Russia}
\affiliation{Institute for Theoretical and Experimental Physics, Moscow, Russia}
\affiliation{Moscow State University, Moscow, Russia}
\affiliation{Institute for High Energy Physics, Protvino, Russia}
\affiliation{Petersburg Nuclear Physics Institute, St. Petersburg, Russia}
\affiliation{Instituci\'{o} Catalana de Recerca i Estudis Avan\c{c}ats (ICREA) and Institut de F\'{i}sica d'Altes Energies (IFAE), Barcelona, Spain}
\affiliation{Uppsala University, Uppsala, Sweden}
\affiliation{Taras Shevchenko National University of Kyiv, Kiev, Ukraine}
\affiliation{Lancaster University, Lancaster LA1 4YB, United Kingdom}
\affiliation{Imperial College London, London SW7 2AZ, United Kingdom}
\affiliation{The University of Manchester, Manchester M13 9PL, United Kingdom}
\affiliation{University of Arizona, Tucson, Arizona 85721, USA}
\affiliation{University of California Riverside, Riverside, California 92521, USA}
\affiliation{Florida State University, Tallahassee, Florida 32306, USA}
\affiliation{Fermi National Accelerator Laboratory, Batavia, Illinois 60510, USA}
\affiliation{University of Illinois at Chicago, Chicago, Illinois 60607, USA}
\affiliation{Northern Illinois University, DeKalb, Illinois 60115, USA}
\affiliation{Northwestern University, Evanston, Illinois 60208, USA}
\affiliation{Indiana University, Bloomington, Indiana 47405, USA}
\affiliation{Purdue University Calumet, Hammond, Indiana 46323, USA}
\affiliation{University of Notre Dame, Notre Dame, Indiana 46556, USA}
\affiliation{Iowa State University, Ames, Iowa 50011, USA}
\affiliation{University of Kansas, Lawrence, Kansas 66045, USA}
\affiliation{Louisiana Tech University, Ruston, Louisiana 71272, USA}
\affiliation{Northeastern University, Boston, Massachusetts 02115, USA}
\affiliation{University of Michigan, Ann Arbor, Michigan 48109, USA}
\affiliation{Michigan State University, East Lansing, Michigan 48824, USA}
\affiliation{University of Mississippi, University, Mississippi 38677, USA}
\affiliation{University of Nebraska, Lincoln, Nebraska 68588, USA}
\affiliation{Rutgers University, Piscataway, New Jersey 08855, USA}
\affiliation{Princeton University, Princeton, New Jersey 08544, USA}
\affiliation{State University of New York, Buffalo, New York 14260, USA}
\affiliation{University of Rochester, Rochester, New York 14627, USA}
\affiliation{State University of New York, Stony Brook, New York 11794, USA}
\affiliation{Brookhaven National Laboratory, Upton, New York 11973, USA}
\affiliation{Langston University, Langston, Oklahoma 73050, USA}
\affiliation{University of Oklahoma, Norman, Oklahoma 73019, USA}
\affiliation{Oklahoma State University, Stillwater, Oklahoma 74078, USA}
\affiliation{Brown University, Providence, Rhode Island 02912, USA}
\affiliation{University of Texas, Arlington, Texas 76019, USA}
\affiliation{Southern Methodist University, Dallas, Texas 75275, USA}
\affiliation{Rice University, Houston, Texas 77005, USA}
\affiliation{University of Virginia, Charlottesville, Virginia 22904, USA}
\affiliation{University of Washington, Seattle, Washington 98195, USA}
\author{V.M.~Abazov} \affiliation{Joint Institute for Nuclear Research, Dubna, Russia}
\author{B.~Abbott} \affiliation{University of Oklahoma, Norman, Oklahoma 73019, USA}
\author{B.S.~Acharya} \affiliation{Tata Institute of Fundamental Research, Mumbai, India}
\author{M.~Adams} \affiliation{University of Illinois at Chicago, Chicago, Illinois 60607, USA}
\author{T.~Adams} \affiliation{Florida State University, Tallahassee, Florida 32306, USA}
\author{J.P.~Agnew} \affiliation{The University of Manchester, Manchester M13 9PL, United Kingdom}
\author{G.D.~Alexeev} \affiliation{Joint Institute for Nuclear Research, Dubna, Russia}
\author{G.~Alkhazov} \affiliation{Petersburg Nuclear Physics Institute, St. Petersburg, Russia}
\author{A.~Alton$^{a}$} \affiliation{University of Michigan, Ann Arbor, Michigan 48109, USA}
\author{A.~Askew} \affiliation{Florida State University, Tallahassee, Florida 32306, USA}
% ...
%
%
\collaboration{The D0 Collaboration\footnote{with visitors from
$^{a}$Augustana College, Sioux Falls, SD, USA,
$^{b}$The University of Liverpool, Liverpool, UK,
$^{c}$DESY, Hamburg, Germany,
$^{d}$Universidad Michoacana de San Nicolas de Hidalgo, Morelia, Mexico
$^{e}$SLAC, Menlo Park, CA, USA,
$^{f}$University College London, London, UK,
$^{g}$Centro de Investigacion en Computacion - IPN, Mexico City, Mexico,
$^{h}$Universidade Estadual Paulista, S\~ao Paulo, Brazil,
$^{i}$Karlsruher Institut f\"ur Technologie (KIT) - Steinbuch Centre for Computing (SCC),
D-76128 Karlsruhe, Germany,
$^{j}$Office of Science, U.S. Department of Energy, Washington, D.C. 20585, USA,
$^{k}$American Association for the Advancement of Science, Washington, D.C. 20005, USA,
$^{l}$Kiev Institute for Nuclear Research, Kiev, Ukraine,
$^{m}$University of Maryland, College Park, Maryland 20742, USA
and
$^{n}$European Orgnaization for Nuclear Research (CERN), Geneva, Switzerland
}} \noaffiliation
\vskip 0.25cm

%% file: Table_peak_1.tex
1&0.000--0.010&13.069~$\pm$~0.052~$\pm$~0.039\\
2&0.010--0.020&12.017~$\pm$~0.049~$\pm$~0.027\\
3&0.020--0.030&10.334~$\pm$~0.046~$\pm$~0.012\\
4&0.030--0.040&~8.652~$\pm$~0.042~$\pm$~0.016\\
5&0.040--0.050&~7.100~$\pm$~0.038~$\pm$~0.008\\
6&0.050--0.060&~5.869~$\pm$~0.034~$\pm$~0.013\\
7&0.060--0.071&~4.863~$\pm$~0.031~$\pm$~0.016\\
8&0.071--0.081&~4.068~$\pm$~0.028~$\pm$~0.007\\
9&0.081--0.093&~3.399~$\pm$~0.024~$\pm$~0.009\\
10&0.093--0.106&~2.803~$\pm$~0.021~$\pm$~0.006\\
11&0.106--0.121&~2.303~$\pm$~0.018~$\pm$~0.006\\
12&0.121--0.139&~1.843~$\pm$~0.014~$\pm$~0.005\\
13&0.139--0.162&~1.442~$\pm$~0.011~$\pm$~0.004\\
14&0.162--0.190&~1.067~$\pm$~0.009~$\pm$~0.003\\
15&0.190--0.227&~0.778~$\pm$~0.007~$\pm$~0.002\\
16&0.227--0.275&~0.524~$\pm$~0.005~$\pm$~0.002\\
17&0.275--0.337&~0.332~$\pm$~0.003~$\pm$~0.002\\
18&0.337--0.418&~0.204~$\pm$~0.002~$\pm$~0.001\\
19&0.418--0.523&~0.115~$\pm$~0.002~$\pm$~0.001\\
\hline
bin&range&$1/\sigma\times(d\sigma/d\phi_{\eta}^{*})(\times 100)$\\
\hline
20&0.523--0.657&~6.428~$\pm$~0.099~$\pm$~0.040\\
21&0.657--0.827&~3.310~$\pm$~0.064~$\pm$~0.029\\
22&0.827--1.041&~1.673~$\pm$~0.041~$\pm$~0.019\\
23&1.041--1.309&~0.818~$\pm$~0.026~$\pm$~0.016\\
24&1.309--1.640&~0.420~$\pm$~0.017~$\pm$~0.010\\
25&1.640--2.049&~0.225~$\pm$~0.011~$\pm$~0.010\\
26&2.049--2.547&~0.120~$\pm$~0.007~$\pm$~0.005\\
27&2.547--3.151&~0.076~$\pm$~0.005~$\pm$~0.004\\
28&3.151--3.878&~0.044~$\pm$~0.004~$\pm$~0.003\\
29&3.878--4.749&~0.026~$\pm$~0.003~$\pm$~0.001\\

%% file: Table_peak_2.tex
1&0.000--0.010&13.404~$\pm$~0.094~$\pm$~0.056\\
2&0.010--0.020&12.189~$\pm$~0.090~$\pm$~0.036\\
3&0.020--0.030&10.635~$\pm$~0.084~$\pm$~0.027\\
4&0.030--0.040&~8.685~$\pm$~0.076~$\pm$~0.030\\
5&0.040--0.050&~7.218~$\pm$~0.069~$\pm$~0.022\\
6&0.050--0.060&~5.836~$\pm$~0.062~$\pm$~0.017\\
7&0.060--0.071&~5.013~$\pm$~0.057~$\pm$~0.027\\
8&0.071--0.081&~4.065~$\pm$~0.050~$\pm$~0.011\\
9&0.081--0.093&~3.382~$\pm$~0.044~$\pm$~0.009\\
10&0.093--0.106&~2.802~$\pm$~0.038~$\pm$~0.010\\
11&0.106--0.121&~2.317~$\pm$~0.032~$\pm$~0.007\\
12&0.121--0.139&~1.827~$\pm$~0.026~$\pm$~0.007\\
13&0.139--0.162&~1.407~$\pm$~0.020~$\pm$~0.008\\
14&0.162--0.190&~1.050~$\pm$~0.016~$\pm$~0.003\\
15&0.190--0.227&~0.764~$\pm$~0.012~$\pm$~0.005\\
16&0.227--0.275&~0.518~$\pm$~0.008~$\pm$~0.002\\
17&0.275--0.337&~0.326~$\pm$~0.006~$\pm$~0.002\\
18&0.337--0.418&~0.194~$\pm$~0.004~$\pm$~0.001\\
19&0.418--0.523&~0.109~$\pm$~0.003~$\pm$~0.001\\
\hline
bin&range&$1/\sigma\times(d\sigma/d\phi_{\eta}^{*})(\times 100)$\\
\hline
20&0.523--0.657&~5.478~$\pm$~0.166~$\pm$~0.050\\
21&0.657--0.827&~2.610~$\pm$~0.102~$\pm$~0.040\\
22&0.827--1.041&~1.167~$\pm$~0.061~$\pm$~0.026\\
23&1.041--1.309&~0.538~$\pm$~0.038~$\pm$~0.017\\
24&1.309--1.640&~0.212~$\pm$~0.022~$\pm$~0.011\\
25&1.640--2.049&~0.104~$\pm$~0.015~$\pm$~0.008\\
26&2.049--2.547&~0.046~$\pm$~0.009~$\pm$~0.005\\
27&2.547--3.151&~0.022~$\pm$~0.006~$\pm$~0.003\\
28&3.151--3.878&~0.013~$\pm$~0.004~$\pm$~0.002\\
29&3.878--4.749&~0.009~$\pm$~0.003~$\pm$~0.001\\

%% file: Table_low_1.tex
1&0.000--0.010&~7.87~$\pm$~0.14~$\pm$~0.12\\
2&0.010--0.020&~7.25~$\pm$~0.13~$\pm$~0.12\\
3&0.020--0.030&~6.98~$\pm$~0.13~$\pm$~0.09\\
4&0.030--0.040&~6.36~$\pm$~0.12~$\pm$~0.06\\
5&0.040--0.051&~5.68~$\pm$~0.11~$\pm$~0.05\\
6&0.051--0.062&~5.15~$\pm$~0.10~$\pm$~0.03\\
7&0.062--0.075&~4.70~$\pm$~0.09~$\pm$~0.03\\
8&0.075--0.092&~3.98~$\pm$~0.08~$\pm$~0.02\\
9&0.092--0.115&~3.21~$\pm$~0.06~$\pm$~0.03\\
10&0.115--0.148&~2.32~$\pm$~0.04~$\pm$~0.03\\
11&0.148--0.198&~1.65~$\pm$~0.03~$\pm$~0.02\\
12&0.198--0.273&~0.98~$\pm$~0.02~$\pm$~0.01\\
\hline
bin&range&$1/\sigma\times(d\sigma/d\phi_{\eta}^{*})(\times 100)$\\
\hline
13&0.273--0.382&52.61~$\pm$~1.07~$\pm$~0.73\\
14&0.382--0.541&25.07~$\pm$~0.63~$\pm$~0.35\\
15&0.541--0.766&11.88~$\pm$~0.36~$\pm$~0.18\\
16&0.766--1.080&~5.05~$\pm$~0.21~$\pm$~0.11\\
17&1.080--1.509&~2.36~$\pm$~0.12~$\pm$~0.07\\
18&1.509--2.087&~1.17~$\pm$~0.08~$\pm$~0.06\\
19&2.087--2.853&~0.40~$\pm$~0.04~$\pm$~0.03\\
20&2.853--3.853&~0.19~$\pm$~0.02~$\pm$~0.02\\

%% file: Table_low_2.tex
1&0.000--0.010&~7.89~$\pm$~0.19~$\pm$~0.12\\
2&0.010--0.020&~7.06~$\pm$~0.18~$\pm$~0.10\\
3&0.020--0.030&~6.77~$\pm$~0.17~$\pm$~0.09\\
4&0.030--0.040&~6.25~$\pm$~0.16~$\pm$~0.05\\
5&0.040--0.051&~5.78~$\pm$~0.16~$\pm$~0.06\\
6&0.051--0.062&~5.37~$\pm$~0.14~$\pm$~0.06\\
7&0.062--0.075&~4.76~$\pm$~0.13~$\pm$~0.02\\
8&0.075--0.092&~4.05~$\pm$~0.10~$\pm$~0.04\\
9&0.092--0.115&~3.18~$\pm$~0.08~$\pm$~0.04\\
10&0.115--0.148&~2.40~$\pm$~0.06~$\pm$~0.02\\
11&0.148--0.198&~1.68~$\pm$~0.04~$\pm$~0.02\\
12&0.198--0.273&~1.00~$\pm$~0.02~$\pm$~0.01\\
\hline
bin&range&$1/\sigma\times(d\sigma/d\phi_{\eta}^{*})(\times 100)$\\
\hline
13&0.273--0.382&54.80~$\pm$~1.48~$\pm$~0.70\\
14&0.382--0.541&26.13~$\pm$~0.85~$\pm$~0.38\\
15&0.541--0.766&11.51~$\pm$~0.49~$\pm$~0.25\\
16&0.766--1.080&~4.39~$\pm$~0.26~$\pm$~0.10\\
17&1.080--1.509&~1.82~$\pm$~0.15~$\pm$~0.06\\
18&1.509--2.087&~0.63~$\pm$~0.07~$\pm$~0.02\\
19&2.087--2.853&~0.26~$\pm$~0.04~$\pm$~0.03\\
20&2.853--3.853&~0.17~$\pm$~0.04~$\pm$~0.02\\

%% file: high/result_0.tex
1 & 0.000--0.010 & 22.48 $\pm$ 1.18 $\pm$ 0.35\\
2 & 0.010--0.020 & 15.34 $\pm$ 0.97 $\pm$ 0.18\\
3 & 0.020--0.030 & 12.73 $\pm$ 0.88 $\pm$ 0.15\\
4 & 0.030--0.040 & ~8.40 $\pm$ 0.72 $\pm$ 0.13\\
5 & 0.040--0.051 & ~8.32 $\pm$ 0.70 $\pm$ 0.11\\
6 & 0.051--0.062 & ~3.87 $\pm$ 0.46 $\pm$ 0.09\\
7 & 0.062--0.075 & ~4.41 $\pm$ 0.45 $\pm$ 0.10\\
8 & 0.075--0.092 & ~3.06 $\pm$ 0.33 $\pm$ 0.10\\
9 & 0.092--0.115 & ~1.65 $\pm$ 0.21 $\pm$ 0.03\\
10 & 0.115--0.148 & ~1.40 $\pm$ 0.16 $\pm$ 0.02\\
11 & 0.148--0.198 & ~0.60 $\pm$ 0.09 $\pm$ 0.02\\
\hline
bin & range & ($1/\sigma$)~$\times$~($d\sigma/d\phistar$) $\times 100$ \\
\hline
12 & 0.198--0.273 & 28.48 $\pm$ 4.98 $\pm$ 0.74\\
13 & 0.273--0.382 & 15.55 $\pm$ 2.98 $\pm$ 0.60\\
14 & 0.382--0.541 & ~6.27 $\pm$ 1.64 $\pm$ 0.33\\
15 & 0.541--0.766 & ~1.50 $\pm$ 0.65 $\pm$ 0.12\\
16 & 0.766--1.080 & ~0.69 $\pm$ 0.39 $\pm$ 0.07\\

%% file: high/result_3.tex
1 & 0.000--0.010 & 28.17 $\pm$ 3.93 $\pm$ 0.57\\
2 & 0.010--0.020 & 22.38 $\pm$ 3.40 $\pm$ 0.34\\
3 & 0.020--0.030 & 18.70 $\pm$ 3.06 $\pm$ 0.41\\
4 & 0.030--0.040 & ~6.61 $\pm$ 1.80 $\pm$ 0.18\\
5 & 0.040--0.051 & ~4.76 $\pm$ 1.48 $\pm$ 0.10\\
6 & 0.051--0.062 & ~3.14 $\pm$ 1.16 $\pm$ 0.12\\
7 & 0.062--0.075 & ~1.91 $\pm$ 0.84 $\pm$ 0.14\\
8 & 0.075--0.092 & ~2.11 $\pm$ 0.78 $\pm$ 0.06\\
9 & 0.092--0.115 & ~1.40 $\pm$ 0.53 $\pm$ 0.06\\
10 & 0.115--0.148 & ~0.68 $\pm$ 0.30 $\pm$ 0.12\\
11 & 0.148--0.198 & ~0.54 $\pm$ 0.22 $\pm$ 0.04\\
\hline
bin & range & ($1/\sigma$)~$\times$~($d\sigma/d\phistar$) $\times 100$ \\
\hline
12 & 0.198--0.273 & ~2.98 $\pm$ 5.47 $\pm$ 0.21\\
13 & 0.273--0.382 & ~7.78 $\pm$ 4.95 $\pm$ 0.47\\
14 & 0.382--0.412 & ~2.24 $\pm$ 2.57 $\pm$ 0.20\\

%% file: phistar_DZero_all_masses.bbl
\begin{thebibliography}{99}
\bibitem{bib:rapidity}
The dilepton rapidity is defined by
$y = (1/2)\ln\left[\left(E+p_z\right)/\left(E-p_z\right)\right],$
where $E$ is the energy and $p_z$ is the momentum component parallel
to the proton beam  direction. The rapidity is given in terms of the momentum
fractions, $x_1$ and $x_2$ of the annihilating quark and antiquark by
$y = (1/2)\ln\left(x_1/x_2\right)$.
\bibitem{bib:thrust}
The thrust axis is defined as a unit vector in the direction of
$\vec{p}_T^{\,\mathrm{(lepton 1)}} - \vec{p}_T^{\,\mathrm{(lepton 2)}}$, where
$\vec{p}_T^{\,\mathrm{(lepton 1)}}$ and  $\vec{p}_T^{\,\mathrm{(lepton 2)}}$ are the lepton momentum vectors in the transverse plane.
\bibitem{bib:pseudo}
Pseudorapidity is defined by 
\mbox{$\eta = -\ln\left[\tan(\theta/2)\right]$},
where $\theta$ is the polar angle with respect to the proton beam direction.
\bibitem{bib:dzero-electrons}
Following Ref.~\protect\cite{bib:dzero-phistar} the momentum of
MC particle-level electrons are defined as the full-momentum vector sum of any photons or electrons 
within a cone of $\Delta R =\sqrt{(\Delta\eta)^2+(\Delta\phi)^2} <
0.2$ around the electron, where $\Delta\eta$ ($\Delta\phi$) is the distance 
in $\eta$ ($\phi$) from the particle-level electron. The dielectron kinematic requirements are:
$70 < M_{\ell\ell} < 110$~GeV, and for both electrons $p_T$~$>$~20~GeV and $|\eta|$~$<$~1.1 or
1.5~$<$~$|\eta|$~$<$~3.
\end{thebibliography}
